\documentclass[11pt]{JHEP3}

\renewcommand{\theequation}{\arabic{section}.\arabic{equation}}

\def\TL{\hfil$\displaystyle{##}$}
\def\TR{$\displaystyle{{}##}$\hfil}

\def\lbldef#1#2{\expandafter\gdef\csname #1\endcsname {#2}}
\def\eqn#1#2{\lbldef{#1}{(\ref{#1})}%
\begin{equation} #2 \label{#1} \end{equation}}
\def\eqalign#1{\vcenter{\openup1\jot
    \halign{\strut\span\TL & \span\TR\cr #1 \cr
   }}}

\def\href#1#2{#2}

\def\[{[}
\def\]{]}

\def\u1{{U(1)}}
\def\ni{\noindent}
\def\ve{\varepsilon}
\def\eps{\epsilon}
\def\lb{\lambda}
\def\N{{\cal N}}

\def\frac#1#2{{#1\over#2}}

\def\half{\frac12}

\def\d{\partial}

\newcommand{\beq}{\begin{equation}}
\newcommand{\eeq}{\end{equation}}
\newcommand{\ber}{\begin{eqnarray}}
\newcommand{\eer}{\end{eqnarray}}

\newcommand{\beqar}{\begin{eqnarray}}
\newcommand{\cN}{{\cal N}}

\newcommand{\eeqar}{\end{eqnarray}}

\title{More on Superstrings in $AdS_3 \times \N$}
\author{Amit Giveon and Ari Pakman   \\ Racah Institute of Physics,
The Hebrew University, Jerusalem 91904, Israel \\
E-mail: \email{giveon@vms.huji.ac.il, pakman@phys.huji.ac.il}}

\abstract{We study superstring theories on  $AdS_3 \times \N$
backgrounds yielding $N=2,3,4$  extended superconformal symmetries
in the dual boundary CFT. In each case the necessary constraints
on the internal worldsheet theory $\N$ are found.}

\preprint{\hepth{0302217}\\RI-02-03}

\begin{document}
\baselineskip=15.5pt
\pagestyle{plain}
\setcounter{page}{1}

\section{Introduction}
The special status of the $AdS_3/CFT_2$ duality in the AdS/CFT context
stems from the fact that
the Virasoro generators of the boundary Conformal
Field Theory (CFT) can be built exactly
from operators in string theory on an $SL(2,R) \times \N$ worldsheet CFT.
This building was done in~\cite{gks,ks},
where it was also shown that an affine algebra in $\N$ can be
uplifted to a similar affine algebra in the
boundary, with a different level.
These constructions are particularly
interesting in the supersymmetric case, where an adequate
field content in $\N$ allows to enlarge the boundary Virasoro
symmetry to  Superconformal Field Theory (SCFT) algebras
with various numbers of supersymmetries.
This was also shown in~\cite{gks},
where the model $SL(2,R)_k \times SU(2)_k \times T^4$
was used to construct a small $N=4$ SCFT algebra
in the boundary theory\footnote{For simplicity,
we discuss only the holomorphic sector of the worldsheet and
the two dimensional spacetime CFT.
The full theory also has a right-handed algebra whose structure
depends on the type of string theory considered: IIA, IIB, heterotic, etc.}
(see \cite{oleg} for a different approach).

Following that work, several backgrounds were explored
yielding other amounts of  boundary supersymmetry.
The extended ($N > 1$) SCFT algebras in two dimension have
been classified in~\cite{Ademollo:1975an,Schoutens:1988ig},
and are $N=2,3$ and two types of $N=4$, small and large.
Their \mbox{R-symmetries} are affine $U(1)$, $SU(2)$,
$SU(2)$ and $SU(2) \times SU(2) \times U(1)$, respectively, with
levels depending on the Virasoro central charge.
In the $AdS_3 \times \N$ backgrounds studied,
the theory $\N$ contains these affine \mbox{R-symmetries},
which are then uplifted to the boundary.

$AdS_3$ backgrounds yielding $N=2$ spacetime SCFT were studied in~\cite{GR,BL}.
For the $N=3$ case, three different backgrounds  were proposed:
\mbox{$\N= SU(3)/U(1)$} and \mbox{$\N=SO(5)/SO(3)$} in~\cite{ags}, and
\mbox{$\N= (SU(2)\times SU(2) \times U(1))/Z_2$} in~\cite{yamaguchi}.
Small $N=4$ was mentioned above.
Finally, for large $N=4$ SCFT, a model was studied in~\cite{Elitzur:1998mm}
with  \mbox{$\N = SU(2) \times SU(2) \times U(1)$},
which  is the minimal field content required for
the boundary \mbox{R-symmetries}.

All these constructions  provide
only {\em sufficient} conditions to yield
boundary extended SCFTs. The purpose of this work is
to study the {\em necessary} constraints
imposed on $\N$ by the existence of each of the
extended SCFT algebras in the boundary theory.

Our results strengthen the relationship between $AdS_3$ boundary
supersymmetry and worldsheet symmetries of the internal CFT $\N$.
This is similar to the case of string compactified to Minkowski space,
where long established results show the intimate connection
between spacetime supersymmetry and the symmetries of the compact
sector~\cite{Banks,Banks:1988yz,Ohta:xi,Shatashvili:zw,Sen:1986mg,Candelas:en}.
In this work we will use CFT techniques similar to those
in~\cite{Banks,Banks:1988yz}.
This makes the results hold for general CFT backgrounds, not necessarily having
a geometrical picture as target spaces of non-linear $\sigma$-models.

Superstring theory on $AdS_3\times\N$
afford boundary SCFT algebras in the NS sector. This
emerges naturally when supercharges are
built from spin fields creating the Ramond sector of the
worldsheet CFT, as for flat space superstrings~\cite{FMS},
and is in accord with expectations
from $AdS_3$ supergravity analysis \cite{Coussaert:jp}.

For $N=2$ and small and large $N=4$, we will show that
the sufficient conditions stated in~\cite{GR,gks,Elitzur:1998mm}, respectively,
are also necessary.
For $N=3$ we  find a set of necessary conditions,
which are an
$SU(2)$-covariant version of
a sufficient condition given in \cite{ags}
to enlarge the $N=2$ SCFT, obtained from
backgrounds of the type \cite{GR,BL}, to $N=3$.

The plan of the work is as follows. Sections 2, 3 and 4 deal
with the $N=2$, $N=3$ and $N=4$ cases, respectively.
In Section 5 we present a short discussion.
Appendix A is devoted to proving some
properties of the R-currents.
In Appendix B, as an illustration, we show
how the results are realized explicitly for
$N=3$ in the background $\N= (SU(2)\times SU(2) \times U(1))/Z_2$.

\section{Spacetime $N=2$ supersymmetry}
In this section we will show that any critical superstring vacuum
of the form $AdS_3 \times \N$ in which
the spacetime theory has (at least) $N=2$ superconformal supersymmetry
must be of the form proposed in~\cite{GR,BL}.
Namely, the internal CFT $\N$ contains an affine $U(1)$ symmetry,
and in the CFT quotient $\N/U(1)$ the worldsheet superconformal symmetry
is extended to $N=2$ supersymmetry.

Let us consider the global part of the $N=2$
spacetime superalgebra in the NS sector:

\begin{eqnarray}
\{Q^+_r,Q^-_s\} & = & 2L_{r+s}+(r-s)R_0\,,           \label{staluno} \\
\[L_m,L_n\] & = & (m-n)L_{m+n}    \,,                \label{staldos} \\
\[L_m,Q^\pm_r\] & = & (\frac{m}2-r)Q^\pm_{m+r} \,,   \label{staltres}\\
\[R_0,Q^\pm_r\] & = & \pm Q^\pm_r    \,,             \label{stalcuatro}
\end{eqnarray}

\def\staluno{(\ref{staluno})}
\def\staldos{(\ref{staldos})}
\def\staltres{(\ref{staltres})}
\def\stalcuatro{(\ref{stalcuatro})}
\ni
with $r,s=\pm\frac12$, $m,n=0,\pm1$, and all other (anti)commutators vanishing.

The above spacetime operators are given
by contour integrals of \mbox{dimension-1} local operators on the
worldsheet. We assume that the global part $L_{0,\pm1}$ of the
Virasoro algebra in spacetime along with the
higher modes $L_n$ are given
by the construction presented in~\cite{gks}, so that,
up to picture-changing \cite{FMS},
\eqn{stvir}{
\eqalign{
L_0 &=-\oint J^3= - \oint e^{-\phi}\psi^3~, \cr
L_{\pm1} & =-\oint (J^1\pm iJ^2)= -\oint e^{-\phi} (\psi^1 \pm i \psi^2)~,
}}

\ni The field $\phi$ is the bosonized superghost and the
superfields $\psi^A+\theta J^A$, $A=1,2,3$, are the affine currents of a
supersymmetric  $SL(2,R)$ WZW model at level $k$. Their OPEs
are\footnote{Indices that go from 1 to 3 will be indicated with
capital letters $(A,B...)$. For those going from 0 to 3 later we will
use lower case letters $(i,j...)$.}:

\eqn{opeads}{
\eqalign{J^A(z) J^B(w) \sim&
{{k\over 2} \eta^{AB} \over (z-w)^2} + {i\epsilon^{ABC}
\eta_{CD} J^D(w) \over z-w}~,\cr
J^A(z) \psi^B(w) \sim &{i\epsilon^{ABC} \eta_{CD} \psi^D(w)
\over z-w}~,\cr
\psi^A(z) \psi^B(w) \sim &{ {k\over 2}\eta^{AB} \over z-w}~,}}
where $\eta^{AB}=(++-)$ and $\epsilon^{123}=1$.
As usual in supersymmetric WZW models,
we can define the currents
\eqn{bosoniccurrent}{
j^A=J^A + \frac{i}{k} \epsilon^{ABC}\eta_{BD}\eta_{CE}\psi^D \psi^E ~,
}
which form an $SL(2,R)$ affine algebra at level $k+2$ and have regular
OPE with the free fermions $\psi^A$.
The central charge of the $AdS_3$ sector is $c=9/2 + 6/k$,
so for a critical theory, $\N$ must
have $c_{\N}=21/2 -6/k$.

$R_0$  is the zero mode of
the U(1) R-current of the $N=2$ algebra in spacetime. The higher modes
$R_n$ can be obtained, say, from the commutators of $R_0$ with $L_n$,
once the latter are introduced.
Alternatively, they can be obtained by the
procedure described in~\cite{gks} to uplift an affine current from
the worldsheet to the boundary spacetime theory (see below).

Let $\psi^0 + \theta J^0$ be the worldsheet supercurrent
corresponding to $R_0$; we have, up to picture-changing,
\eqn{stul}{R_0= \sqrt{2k}\oint  J^0= \sqrt{2k}\oint  e^{-\phi} \psi^0~,}
where $J^0$ and $\psi^0$ are orthogonal and canonically normalized
(see Appendix A):
\eqn{jj}{
\eqalign{
J^0(z) J^0(w) & \sim {1 \over (z-w)^2}~, \cr
\psi^0(z) \psi^0(w) & \sim \frac1{z-w}~,\cr
J^0(z)\psi^0(w) & \sim 0~.
}}
To see that the choice \jj\ leads to the normalization of \stul, recall
\cite{gks} that the higher modes of the R-current have the form
\eqn{rmodos}{
R_n= a \oint  J^0 \gamma^n~,
}
\ni
and they satisfy
\eqn{conmumodosu1}{
\[R_m,R_n\]=a^2pm\delta_{n+m,0}~,
} \ni
with \eqn{pdef}{ p=\oint  {\partial_z \gamma \over
\gamma}~, }
where $\gamma$ is the zero-dimension field of the
$(\beta,\gamma)$ pair appearing in the Wakimoto free-field representation
of the algebra \opeads. We want to show that $a=\sqrt{2k}$. Indeed,
consistency of the $N=2$ algebra \cite{Ademollo:1975an,sw}
in spacetime implies $a^2p=c_{st}/3$, where $c_{st}$ is the
central charge of the Virasoro algebra in the dual boundary CFT,
given by  $c_{st}=6kp$.~\footnote{The reader is
referred to \cite{gks} for details of the construction.}
This fixes $a=\pm\sqrt{2k}$.~\footnote{In the
following we will consider only the expressions for
$a=+\sqrt{2k}$, but everything holds for the other choice,
changing signs appropriately.}.

As for the relation between this $U(1)$ current and the three
$SL(2,R)$ currents, in Appendix A we  show that the commutation relations in
spacetime \staluno\ -- \stalcuatro\ and the fact that the worldsheet theory is
supersymmetric force $\psi^0 + \theta J^0$ to lie {\it entirely}
in the internal CFT $\N$.

\subsection{Properties of the spacetime supercharges}
Regarding the four spacetime supercharges  $Q^\pm_r$,
we will only assume that,
as for superstrings in flat space~\cite{FMS}, they are obtained
from operators that create the worldsheet Ramond sector:
\eqn{gese}{
Q^\pm_r=b^{\pm}_r\oint  e^{-\frac\phi2} S^\pm_r~,\qquad
r=\pm\frac12~,}
where $S^\pm_r$ are spin fields \cite{fqs,FMS} and $b^{\pm}_r$ are constants.
Since unbroken worldsheet supersymmetry requires $G_0^2=L_0^{ws}-c/24=0$
on the Ramond ground state, we have
$\Delta(S^\pm_r)=\frac{c}{24}=\frac58$. This is compatible with the on-shell
condition $\Delta(Q)=0$ for the $Q$'s in \gese.
In the presence of spin fields,
the fermionic parts of the superfields become double-valued, i.e.,
integer modded on the plane, and
the bosonic fields remain single-valued.


In our case, we have identified four free fermions $\psi^i$, $i=0,1,2,3$,
which are the lower components of superfields.
Since, at the moment, we are not given further data on the field content
of the worldsheet theory $\N$,
we only know that the whole set of spin fields is in a representation
of the algebra satisfied by the zero modes of $\psi^i$.
This is the four dimensional Clifford algebra:
\eqn{cliff}{
 \{ \psi^i_0,\psi^j_0 \} = g^{ij}{k^i}~,
}
with $g^{ij}=(+,+,+,-)$, $k^0=1$ and $k^A=k/2$. In particular,
we can decompose the whole set of spin fields
into irreducible representations (irreps) of \cliff.
By a result proved in~\cite{Pauli},
all the irreps of \cliff\ have dimension 4 and are all equivalent.
Hence the OPE of the spin fields with the four fermions is
\cite{fqs,Cohn:1986bn}:
\eqn{propuno}{
\psi^i(z)S_{\ve_1,\ve_2,\lb}(w) \sim {
(\psi^i_0)^{\ve'_1,\ve'_2}_{\ve_1,\ve_2}
S_{\ve'_1,\ve'_2,\lb}(z) \over (z-w)^{\frac12} }~, }
where $\lb$ is an index indicating to which
particular irrep we refer, and $\ve_{1,2}=\pm 1/2$.
The spacetime algebra \staluno\ -- \stalcuatro\ implies that
$S^\pm_r$ in \gese\ form such an irrep
(this will be described explicitely below).

By Wick rotating $\psi^3$, these irreps realize
(anti)spinorial representations of the \mbox{level-2} $SO(4)_2$ affine
algebra constructed out of bilinears of $\psi^i(z)$. Since the
$SO(4)_2$ currents are bosonic dimension-1 fields, their OPEs with
the spin fields are single-valued. For such (anti)spinorial
representations of $SO(4)_2$, the weights are $\pm \half$.
Choosing for the Cartan subalgebra the fields
\eqn{haches}{
\d H_1 = \frac2k \psi^1 \psi^2~, \qquad
\d H_2  = -i\sqrt{\frac2k} \psi^0 \psi^3~, }
with
\eqn{hachehache}{ H_I(z)H_J(w) \sim - \delta_{IJ} \log(z-w)~, \qquad
I,J=1,2~, }
and $H_1^\dagger=H_1$, $H_2^\dagger=-H_2$, we must have
\eqn{aut}{ i\d H_{1,2}(z) S_{\ve_1,\ve_2,\lb}(w) \sim
{\ve_{1,2}} {S_{\ve_1,\ve_2,\lb}(w) \over z-w} \,, \qquad
\ve_{1,2}= \pm \half ~. }
It follows that the  fields
$S_{\ve_1,\ve_2,\lb}(z)$ can be written as
\eqn{ese}{
S_{\ve_1,\ve_2,\lb}= e^{i {\ve_1}H_1+ i{\ve_2} H_2} e^{i \pi \ve_2
N_1} ~ \tilde{\Sigma}_{\lb}~, }
where we have written explicitly the
cocycle $e^{i \pi \ve_2 N_1}$ \cite{kostelecky}, which is
necessary for \ese\ to yield a good representation of \propuno.
The number operator $N_1$ is given by
\eqn{numberop}{ N_1 =i\oint
 \partial H_1(z)~. }

\ni
Equations \aut -\ese\ imply that
\eqn{hachesigma}{
H_I(z)\tilde{\Sigma}_{\lb}(w) \sim 0~,
}
and from this we have in turn\footnote{
For example, $\psi^0(z)\tilde{\Sigma}_{\lb}(w)
\sim {(e^{iH_2(z)}-e^{-iH_2(z)})
\over i \sqrt{2}} \tilde{\Sigma}_{\lb}(w) \sim 0$, and so on.}
\eqn{nhr}{
\psi^i(z)\tilde{\Sigma}_{\lb}(w)  \sim 0 ~.}

\ni
For future reference we remind that
{\setlength\arraycolsep{2pt}
\renewcommand{\arraystretch}{1.8}
\begin{equation}
\label{hachexp}
\begin{array}{rcrcl}
e^{\pm iH_1}
& = & \frac1{\sqrt{k}}  (\psi^1 & \mp & i \psi^2)~,  \\
e^{\pm i\pi N_1} e^{\pm iH_2}  & = &
{1\over \sqrt{2}} \psi^0 & \mp & {1 \over \sqrt{k}}\psi^3 ~.
\end{array}
\end{equation}}
\def\hachexp{(\ref{hachexp})}



\ni
Note that in \ese\ we have obtained for $S_{\ve_1,\ve_2,\lb}$
the structure usually assumed, with
$\tilde{\Sigma}_{\lb}$ generally being a spin field for the other fermionic
fields present in the theory. But the path we have taken is meant to stress
that relations \ese, \hachesigma\ and \nhr\
hold under the general property of $S_{\ve_1,\ve_2,\lb}$ being
a representation of the algebra of the four fermions $\psi^i$,
regardless of the further structure of the $\N$ CFT.

According to \staltres,
the charge of $S^{\pm}_r
~(r=\pm \frac12)$ under $L_0$ is $-r$. Since $e^{i {\ve_1}H_1}$
and $e^{i {\ve_2}H_2}$ have charges $\ve_1$ and $0$ under $L_0$,
respectively, we conclude that the identification $r=-\ve_1$
should be made, and that $\tilde{\Sigma}_{\lb}$ is uncharged under
$L_0$. This is consistent with the action of $L_{+1}$ ($L_{-1}$),
which lowers (raises) the eigenvalue of $L_0$ by one, provided
that $\tilde{\Sigma}_{\lb}$ is untouched by $J^{\pm}$. It follows
that\footnote{An $L_0$ charge of $-r$ for $S_{\ve_1,\ve_2,\lb}$
could also have been obtained by choosing $\ve_1=r$ and letting
$\tilde{\Sigma}_{ \lb}$ carry charge $-2r$ under $L_0$, but this
option is inconsistent with the action of $L_{\pm 1}$.}
\eqn{slsigma}{ J^A(z) \tilde{\Sigma}_{\lb}(w) \sim 0~. }

\ni
Having identified the $L_0$-charge of $S_{\ve_1, \ve_2,\lb}$ as
$\ve_1$, we expect $S^{\pm}_r$ to be obtained from
$S_{-r,\ve_2,\lb}$. Imposing  that the supercharges $Q^{\pm}_r$ of
\gese\ satisfy  \staluno, along with
$\{Q^\pm_r,Q^\pm_s\}=0$, fixes\footnote{We omit the cocycles from
now on.}
\eqn{defmas}{ S^\pm_r =e^{-i r(H_1 \mp
H_2)}\tilde{\Sigma}^\pm \,,\,\,\,\,\,~~~~~r=\pm \frac12~, } where we
have relabelled $\tilde{\Sigma}_{\lb}$ accordingly. Consistency of
the algebra \staluno\ -- \stalcuatro\ requires
\eqn{consis}{ \eqalign{ \tilde{\Sigma}^+(z)
\tilde{\Sigma}^-(w)  & \sim  {1 \over (z-w)^{3 \over 4}}\,, \cr
\tilde{\Sigma}^{\pm}(z) \tilde{\Sigma}^{\pm}(w)  & \sim  O(w)
(z-w)^{3 \over 4}~. }}
Finally, the constants $b^{\pm}_r$ in
\gese\ are determined to be $(4k)^{\frac14}$ up to phases.

According to \stalcuatro\ and \stul,  the operators $S^\pm_r$
are charged under $J^0$ with charges $\pm {1 \over \sqrt{2k}}$.
But from  the results of Appendix A,
we know that the operators $e^{-i r(H_1 \mp
i H_2)}$ are neutral under $J^0$, so the charges are in
$\tilde{\Sigma}^\pm_r$.
Writing \eqn{jzeta}{ J_0=i \d Y\,,} with
\eqn{nory}{Y(z)Y(w) \sim -\log(z-w)~,}
we must have
\eqn{sepa}{
\tilde{\Sigma}^\pm= e^{\pm i\sqrt{\frac1{2k}}Y}\Sigma^\pm~,}
and
\eqn{ynsi}{
Y(z)\Sigma^\pm(w) \sim 0 ~.}

\ni Collecting our results until now, we have
\eqn{colect}{
S^\pm_r= e^{-ir(H_1 \mp H_2)}e^{\pm i \sqrt{\frac1{2k}}Y}
\Sigma^\pm~, }
and from \nhr, \slsigma\ and \ynsi\ it follows that
$\Sigma^\pm$ belong entirely to $\N/U(1)$, i.e.,
\eqn{opejfi}{
J^i(z) \Sigma^\pm(w) \sim \psi^i(z) \Sigma^\pm(w) \sim 0~. }
Moreover,
\eqn{defbeta}{ \Delta(\Sigma^\pm) \equiv  {\beta^2 \over
2}={3 \over 8} - {1 \over 4k} = {c_{\N/U(1)} \over 24}~, }
as should
be for operators that create the Ramond sector ground state
of $\N/U(1)$ with unbroken supersymmetry.

After the introduction of the $e^{\pm i\sqrt{\frac1{2k}}Y}$
factors in $\tilde{\Sigma}^\pm$, eq. \consis\ turns into
\eqn{opesigma}{
\eqalign{ \Sigma^-(z) \Sigma^+(w) \sim & {1 \over (z-w)^{\beta^2}}~, \cr
\Sigma^\pm(z) \Sigma^\pm(w)
\sim & (z-w)^{\beta^2}  O^\pm(w)~,
}}
where $\Delta(O^\pm)=\frac32-\frac1k$.
We can now use standard techniques \cite{Banks,Polchinski}.
Consider the four-point function
\eqn{fourpoint}
{
f(z_j)= \langle \Sigma^-(z_1) \Sigma^+(z_2) \Sigma^-(z_3)
\Sigma^+(z_4) \rangle~.
}
Using \defbeta\ and $SL(2,C)$ invariance, it can be written as
\eqn{fpd}{
f(z_j)= \left( \frac{z_{13}z_{24}}{z_{12}z_{34}z_{14}z_{23}}
\right)^{\beta^2} \tilde{f}(x)~,
}
where $z_{jk}=z_j-z_k$ and $x\equiv
\frac{z_{12}z_{34}}{z_{13}z_{24}}$. As the points $z_i$
coincide pairwise, the OPEs \opesigma\
imply that $\tilde{f}$ is an analytic function and is bounded
for $x \rightarrow \infty$, hence it is a constant.
Expanding as $z_{12} \rightarrow 0$, \fpd\ becomes
\eqn{fpdl}{
f(z_j)=z_{12}^{-\beta^2} z_{34}^{-\beta^2} \left( 1 + \beta^2
\frac{z_{12}z_{43}}{z_{23}z_{24}} \right) ~,
}
where the first term and \opesigma\ fix $\tilde{f}$ to $1$.
The presence of a second term in \fpdl\ implies that in the
$\Sigma^- \Sigma^+$ OPE expansion there is a dimension-$1$ field
$M$,
\eqn{opesigmados}{
\Sigma^-(z_1)\Sigma^+(z_2)  \sim {1 \over z_{12}^{\beta^2}}
\left[ 1 + z_{12}{M(z_2)\over 2} \right]~,
}
whose three-point function with $\Sigma^- \Sigma^+$ is determined
from \fpdl\ to be
\eqn{threepoint}{
\langle M(z_2) \Sigma^-(z_3) \Sigma^+(z_4) \rangle = 2 \beta^2
z_{34}^{1-\beta^2}z_{23}^{-1}z_{24}^{-1}~.
}
Taking the limits $z_{34},z_{23},z_{24} \rightarrow 0$, we obtain
\eqn{normaeme}{ M(z)M(w)
\sim {4 \beta^2 \over (z-w)^2}~, }
and
\eqn{jota}{ M(z)\Sigma^\pm(w)
\sim \pm{2 \beta^2 \Sigma^\pm \over z-w} ~. }
Defining\footnote{The sign choice in \jota\ is arbitrary.}
\eqn{zeta}{M \equiv 2 i\beta \d Z~,}
with
\eqn{zetazeta}{Z(z)Z(w) \sim -\log(z-w)~,}
we have
\eqn{sigmazetapi}{\Sigma^\pm = e^{\pm i \beta Z} \Pi^\pm\,,}
with
\eqn{zetapi}{Z(z)\Pi^\pm(w) \sim 0\,.}
\ni
But since $\Delta(\Sigma^\pm)=
{\beta^2 \over 2}$, it follows that $\Delta(\Pi^\pm)=0$, hence $\Pi^\pm =1$.
Summing up, the spin fields have the form
\eqn{spinfield}{
S^\pm_r= e^{-ir(H_1 \mp H_2)}e^{\pm i \sqrt{\frac1{2k}}Y} e^{\pm i \beta Z}~.}

\subsection{Worldsheet symmetries}
Since the spacetime theory is supersymmetric,
the supercharges $Q^{\pm}_r$ take physical states
into physical states, hence they should commute with the BRST operator.
This means that in the OPEs between the worldsheet supercurrent $G$ and
$S^\pm_r$ no
$(z-w)^{-\frac32}$ terms appear. The supercurrent is given by
\eqn{getotal}{
G=G_{AdS_3} + G_{U(1)} + G_{\N/U(1)}~,}
where
\eqn{gterm}{
\eqalign{
G_{AdS_3} & = \frac2k(\psi^Aj_A+ \frac{2i}k \psi^1\psi^2\psi^3)~,\cr
G_{U(1)} & = J^0\psi^0~,
}}
and by definition
\eqn{bydef}{
J^iG_{\N/U(1)} \sim \psi^iG_{\N/U(1)} \sim 0~.}

\ni
In the computation of the OPEs between  $G$ and $S^\pm_r$,
eqs. \opejfi\ and \bydef\ imply that singular terms come only from
\eqn{relev}{
( G_{AdS_3} + G_{U(1)})(z)\,\,
e^{-ir(H_1 \mp H_2)}e^{\pm i\sqrt{\frac1{2k}}Y}(w)~,
}
and
\eqn{relevdos}{
G_{\N/U(1)}(z)\,\,e^{(\pm i \beta Z)}(w)~.
}
In the OPEs \relev\ two $(z-w)^{-\frac32}$ terms appear which
cancel each other. For the computation it is convenient to express
(see \haches, \hachexp)
\eqn{conviene}{\eqalign{
\frac{4i}{k^2} \psi^1\psi^2\psi^3 & =
\frac{i}{\sqrt{k}} \d H_1 (e^{-iH_2}-e^{iH_2})\,,\cr
J^0\psi^0 & = \frac1{\sqrt{2}} J^0 (e^{-iH_2}+e^{iH_2})~.
}}
Now, given the $U(1)$ current $M$ in \zeta,
every operator $\Phi$ in the theory can be decomposed into terms
with definite $M$-charge $q$ as
\eqn{pdec}{
\Phi=\sum_q :e^{{iq \over 2 \beta}Z}P_q(M):\tilde{\Phi}_q \,,
}

\ni
where $P_q(M)$ is a polynomial in $M(z)$ and its derivatives,
and $\tilde{\Phi}_qM \sim 0$.

The absence of $(z-w)^{-\frac32}$ terms in
\relevdos\ and dimensional analysis imply that the only
terms allowed when expressing $G_{\N/U(1)}$ as \pdec\ are
\eqn{gsd}{
\eqalign{
G_{\N/U(1)} & = G_{\N/U(1)}^+ + G_{\N/U(1)}^- ~,\cr
G_{\N/U(1)}^{\pm} & = \tau^{\pm} e^{\pm{ i \over 2 \beta}Z}~,
}}
with
\eqn{susydos}{
\eqalign{
M(z)G_{\N/U(1)}^{\pm}(w) & \sim \pm \frac{G_{\N/U(1)}^{\pm}(w)}{z-w}~,
\cr
Z(z) \tau^\pm(w) & \sim 0~.
}}
From this it follows  that $\N/U(1)$ has $N=2$ supersymmetry,
with $M$ being the $U(1)$ \mbox{R-current}. The rest of the $N=2$ commutators
can be obtained as in~\cite{Banks}, using Jacobi identities.
Note from \defbeta\ and \normaeme\ that $M$
has the correct normalization for
the $N=2$ algebra, namely, $M(z)M(w)\sim {c_{\N/U(1)} \over 3}(z-w)^{-2}$.

\section{Spacetime $N=3$ supersymmetry}
\setcounter{equation}{0}

Let us consider now the case of spacetime $N=3$ supersymmetry
in the NS sector. The global subalgebra is \cite{Ademollo:1975an}
\begin{eqnarray}
\{Q_r^a, Q_s^b\} & = & 2 \delta^{a,b}~,
\, L_{r+s} +i \eps^{ab}_{~~c} (r-s) T^c_{r+s}~, \label{uno} \\
\[L_m, L_n\] & = & (m-n) L_{m+n}~, \label{dos} \\
\[T^a_0, T^b_0\] & = & i \eps^{ab}_{~~c} \, T^c_0~, \label{tres} \\
\[L_m, Q_r^a \] & = & \left(\half m - r\right) Q^a_{m+r}~, \label{cuatro}\\
\[T^a_0, Q_r^b\] & = & i \eps^{ab}_{~~c} Q_r^c~, \label{cinco}
\end{eqnarray}
where $m,n=0, \pm1$, $a,b,c=1,2,3$ and $r,s=\pm\half$,
and all other (anti)commutators vanish.
All these spacetime operators are again obtained by contour
integrals of \mbox{dimension-1} local fields on the worldsheet.
The three operators $L_{0,\pm1}$ are given again
by \stvir -\opeads.

The three operators $T^a_0$ are the zero modes of the $SU(2)$
R-current of the $N=3$ algebra in spacetime.
Let $\chi^a + \theta K^a$ be the three dimension-$\half$ worldsheet
supercurrents corresponding to $T^a_0$; we have, up to picture-changing,
\eqn{stsutwo}{T^a_0=  \oint  K^a=  \oint  e^{-\phi} \chi^a~.}

\ni
The supercurrents $\chi^a + \theta K^a$ form an affine $SU(2)$
superalgebra in $\N$,
which is uplifted to an affine $SU(2)$ algebra in the dual boundary CFT,
according to the construction of \cite{gks}.
The worldsheet level $k'$ of this $SU(2)$ affine algebra
is again fixed by looking at the higher modes of
the spacetime $SU(2)$ affine currents, which are given by
\eqn{su2modos}{
T^a_n=  \oint  K^a \gamma^n~,
}
and which satisfy
\eqn{cmsu2}{
[T^a_m,T^b_n]= i\eps^{ab}_{~~c}T^c_{m+n} +
\frac{k_{st}}2 \, m\,\delta_{a,b}\,\delta_{n+m,0}~,
}
where the spacetime $SU(2)$ level is $k_{st}=k'p$ \cite{gks},
with $p$ given by \pdef.
Consistency of the $N=3$ algebra (see \cite{Ademollo:1975an,sw})
in spacetime implies
$k_{st} = {2\over 3}c_{st}$, which is equivalent to $k'p={2\over 3}6kp$,
and hence $k'=4k$.
Note that the level-dependent normalization of the $T^a_0$
is the same as that
of the $R_0$ operator in the $N=2$ case (see eq. \stul).
We will use these facts below.

We shall also use the purely bosonic $\sigma$-model
and purely fermionic contributions to the total currents:
\eqn{boscurdos}{\eqalign{
k^a & =  K^a - k^a_{su(2)_2}~, \cr
k^a_{su(2)_2}& =- \frac{i}{4k} \epsilon^{a}_{~bc}\chi^b \chi^c~.
}}
The currents $k^a$ and $k^a_{su(2)_2}$ are two commuting affine $SU(2)$
currents, at levels $4k-2$ and $2$, respectively,
and the bosonic $k^a$ commute with the fermions $\chi^a$.

\subsection{Properties of the spacetime supercharges}
Again, the supercharges are given by
\eqn{gese2}{
Q^a_r=b^{a}_r\oint
e^{-\frac\phi2} S^a_r~,\qquad r=\pm\frac12~.}

\ni
In the Ramond sector, the algebra of the fermionic zero modes is
now
\eqn{cliffdos}{\eqalign{
\{ \psi^A_0,\psi^B_0 \} & = \eta^{AB}k/2 ~,
\cr
\{ \chi^a_0,\chi^b_0 \} & = \delta^{a,b}2k ~,
\cr
\{ \psi^A_0,\chi^a_0 \} & =0~.
}}
For the $SO(6)_2$ affine algebra obtained by Wick rotating $\psi^3$, we choose
the Cartan subalgebra given by
\eqn{hachesdos}{
\d H_1  = \frac2k \psi^1 \psi^2~, \qquad
\d H_2  = -\frac{i}k \chi^3 \psi^3~, \qquad
\d H_3  = \frac1{2k} \chi^1 \chi^2~,
}
with
\eqn{hachehachedos}{ H_I(z)H_J(w) \sim - \delta_{IJ} \log(z-w)\,,\qquad
I,J=1,2,3~, }
and $H_{1,3}^\dagger=H_{1,3}$, $H_2^\dagger=-H_2$.

As in the $N=2$ case, there is a basis for the spin fields
of the form
\eqn{esedos}{
S_{\ve_1,\ve_2,\ve_3, \lb}=
e^{i {\ve_1}H_1+ i{\ve_2} (H_2 + \pi N_1)+
i{\ve_3}(H_3+ \pi N_2+ \pi N_1)} \Lambda_{\lb}~,
}
with
\eqn{nhrdos}{
\psi^A(z)\Lambda_{\lb}(w)  \sim 0~, \qquad \chi^a(z)\Lambda_{\lb}(w)  \sim 0~,}
and
\eqn{numberopdos}{
N_{1,2} =i\oint \partial H_{1,2}(z) \,.
}

\ni
The fields \esedos\ provide a representation of \cliffdos\ (as in \propuno),
with any other representation given by a linear change of basis of the spin
fields. For future
reference we remind that
{\setlength\arraycolsep{2pt}
\renewcommand{\arraystretch}{1.6}
\begin{equation}
\label{hachexp2}
\begin{array}{rcrcl}
e^{\pm iH_1} & = & \frac1{\sqrt{k}}  (\psi^1 & \mp & i \psi^2)~,  \\
e^{\pm i(H_2 + \pi N_1)} & = & {1\over 2\sqrt{k}} \chi^3 & \mp &
{1 \over \sqrt{k}}\psi^3~, \\
e^{\pm i(H_3 + \pi N_2 + \pi N_1)} & = & \frac1{2\sqrt{k}}
(\chi^1 & \mp & i \chi^2)~,
\end{array}
\end{equation}}
and from \boscurdos, we have
\eqn{bosfer}{\eqalign{
k^3_{su(2)_2} & = -i \d H_3~, \cr
k^{\pm}_{su(2)_2} & = k^1_{su(2)_2} \pm i k^2_{su(2)_2} =
\pm (e^{-i H_2 }+ e^{+i H_2 }) \, e^{\mp i H_3 }~.
}}

\ni
Now, the algebra (\ref{uno}) -- (\ref{cinco}) has three
$N=2$ subalgebras, given by, say,
\eqn{enedos}{\eqalign{
Q^{\pm}_r & = \frac1{\sqrt{2}}(Q_r^1 \pm i Q_r^2)\,,\cr
R_0 & = T_0^3 \,.
}}
The other two $N=2$ subalgebras are obtained
by cyclic permutations.
From the result of the previous section we know that for the
representation (\ref{esedos}), the spin fields
corresponding to the supercharges $Q^{\pm}_r$ in \enedos\ must
be those obtained from identifying $\ve_1= -r$ and
$\ve_2=\pm r$ in \esedos, and that
\eqn{jotalam}{
J^A(z)\Lambda_{\lb}(w) \sim 0~.
}

\ni
The operators $\tilde{\Sigma}^{\pm}$  of \defmas,
having charge $\pm 1$ under $T^3_0$ and
satisfying \consis,
must be obtained from
$e^{i\ve_3 H_3} \Lambda_{\lb}$.
Since
$e^{i \ve_3 H_3}$
has charge $-\ve_3 = \pm \half$ under $T^3_0$,
the other $\pm \half$  charge
must be carried by\footnote{The possibility that $e^{i \ve_3 H_3}$
carries charge $\mp \half$ and $\Lambda_{\lb}$ carries charge $\pm \frac32$
must be discarded because it is inconsistent with the action of $T^{\pm}_0$.}
$\Lambda_{\lb}$.
We conclude that
there are two fields $\Lambda^{\pm}$,
which from \nhrdos\ are charged under $k^3$ (see \boscurdos) as
\eqn{cargalambda}{
k^3(z) \Lambda^{\pm} (w) \sim {\pm \half \Lambda^{\pm}(w) \over z-w} \,,
}
and that we should make the following identification
(we omit the cocycles from now on):
\eqn{iden}{
\tilde{\Sigma}^{\pm} = e^{\mp \frac{i}2 H_3}\Lambda^{\pm}.
}
Then from \consis\ we obtain
\eqn{opelambda}{
\eqalign{
\Lambda^+(z) \Lambda^-(w)  & \sim  {1 \over (z-w)^{\frac12}}~, \cr
\Lambda^{\pm}(z) \Lambda^{\pm}(w)  & \sim  O(w) (z-w)^{\frac12}~.
 }}
A similar analysis to the one following \opesigma, shows
that in the first OPE \opelambda\
there exists a dimension-1 field $i\d X^3$:
\eqn{opelcom}{
\Lambda^+(z) \Lambda^-(w)   \sim  {1 \over (z-w)^{\frac12}}
\left[1 - (z-w)\frac{i \d X^3(w)}{\sqrt{2}} \right]~,
}
with
\eqn{equistres}{X^3(z)X^3(w) \sim -\log (z-w)~,}
such that
\eqn{lambdaequis}{
\Lambda^{\pm}= e^{\mp \frac{i}{\sqrt{2}}X^3} \,.
}
{}From \cargalambda\ and \opelambda\ we find the three-point function
\eqn{trespuntos}{
\langle k^3(z_1) \Lambda^+(z_2) \Lambda^-(z_3) \rangle =
\half z_{12}^{-1}\, z_{13}^{-1}\, z_{23}^{\half}~,
}
and taking $z_{23}\rightarrow 0$, from \opelcom\ we have
\eqn{cargaequis}{
k^3(z) i \d X^3(w) \sim  {- 1 \over \sqrt{2} \, (z-w)^2} \,.
}
The above expression allows to write $K^a$ as (see \boscurdos)
\eqn{separoequis}{\eqalign{
K^3 & =  \hat{k}^3 +k^3_{su(2)_1} +k^3_{su(2)_2} ~,\cr
K^{\pm} & = K^1 \pm i K^2 = \hat{k}^{\pm} +
k^{\pm}_{su(2)_1} + k^{\pm}_{su(2)_2}~,
}}
where
\eqn{jsuone}{k^3_{su(2)_1}\equiv -{i\over\sqrt{2}}\d X^3~, \qquad
k^{\pm}_{su(2)_1}\equiv k^1_{su(2)_1}\pm ik^2_{su(2)_1}=e^{\mp i\sqrt{2}X^3}~.}
\ni
Note that the $SU(2)_{4k}$ currents $K^a$ are decomposed
into three affine $SU(2)$ currents $\hat{k}^a$,
$k^a_{su(2)_1}$ and $k^a_{su(2)_2}$, with levels
$4k-3$, $1$ and $2$, respectively.
The $SU(2)_1$ is the theory of a single compact scalar $X^3$, and the
$SU(2)_2$ is the theory of three free fermions whose currents are given in
\boscurdos, \bosfer.
The three sets of $SU(2)$ currents commute among themselves because
\eqn{katecho}{
\hat{k}^a(z) \chi^b(w) \sim  \hat{k}^a(z) \d X^3(w)
\sim \d X^3(z) \chi^a(w) \sim 0~.
}

\ni
Summing up, from \iden\ and \lambdaequis\ we find that the spin fields
corresponding to the supercharges in \enedos\ are
\eqn{}{
S^{\pm}_r = e^{-ir (H_1\mp H_2) \mp {i\over 2} H_3}
e^{\mp \frac{i}{\sqrt{2}}X^3}~,
}
and using the decomposition \separoequis,
the operators $S^3_r$ can now be obtained by applying $K^{\pm}$
to $S^{\pm}_r$. Defining
\eqn{defin}{
S_{[\pm,\pm,\pm,\pm]}=
e^{\frac{i}2 ( \pm H_1  \pm H_2   \pm H_3) \pm \frac{i}{\sqrt{2}}X^3}\,,
}
the whole algebra is then generated by the following spin fields:
{\setlength\arraycolsep{1pt}
\renewcommand{\arraystretch}{1.6}
\begin{equation}
\label{spintodos}
\begin{array}{lcl}
S^-_{\half}  & = &            S_{[-,-,+,+]}~, \\
S^3_{\half}  & = & \frac12(   S_{[-,-,+,-]} + i S_{[-,+,-,+]})~,\\
S^+_{\half}  & = &  -i        S_{[-,+,-,-]}~, \\
S^-_{-\half} & = &            S_{[+,+,+,+]}~, \\
S^3_{-\half} & = & \frac12 (  S_{[+,+,+,-]} + i S_{[+,-,-,+]})~, \\
S^+_{-\half} & = & -i         S_{[+,-,-,-]}~.
\end{array}
\end{equation}}
\def\spintodos{(\ref{spintodos})}
\ni Note that the $e^{\frac{i}2 (\pm H_1\pm H_2\pm H_3)}$ fields provide
(several) spin-$\mathbf{\half}$ representations of the $SU(2)_2$ algebra
of the $\chi^a$.
The fields $\Lambda^{\pm}$
in \lambdaequis\ are in turn two primaries of a spin-$\mathbf{\half}$
representation of the $SU(2)_1$ made out of $X^3$.
The spin-$\mathbf{1}$ representation
$S^a_r$ emerges then as the symmetric part of the
$\mathbf{\half} \otimes \mathbf{\half}$ product of these two
spin-$\mathbf{\half}$ representations.


\subsection{Worldsheet symmetries}

Let us explore now  the consequences induced by the structure of the spin
fields on the properties of the worldsheet theory.
The BRST condition again implies that no
$(z-w)^{-\frac32}$ singularities appear in the OPEs between $G$
and the spin fields \spintodos. The worldsheet supercurrent $G$ is
\eqn{getotaldos}{ G  = G_{AdS_3} + G_{SU(2)} + G_{\N/SU(2)}\,, }
with $G_{AdS_3}$ given by \gterm\ and
\eqn{gesudos}{ G_{SU(2)}=
\frac1{2k} \left(\chi^a k_a - {i\over 2k}\chi^1 \chi^2 \chi^3\right) \,. }
The terms whose OPE with
the spin fields might give a $(z-w)^{-\frac32}$ singularity can be
written as
\eqn{tresmedios}{ \eqalign{
G= \ldots & +
\frac{i}{\sqrt{k}} \d H_1 \left( e^{-iH_2} - e^{+iH_2} \right)  +
\frac1{2\sqrt{k}} \left( e^{-i \sqrt{2}X^3 + iH_3} + e^{i \sqrt{2}X^3 -
iH_3} \right) \cr
 & -\frac1{2\sqrt{k}}\left( \frac{i}{\sqrt{2}} \d X^3 + i\d H_3 \right)
\left(e^{-iH_2} + e^{+iH_2}\right) + G_{\N/SU(2)}~,
}}
where the ``dots'' stand for terms that manifestly do not contribute
$(z-w)^{-\frac32}$ singularities.
It can be checked that all the $(z-w)^{-\frac32}$
terms in the OPEs cancel among themselves
for the first three terms in \tresmedios. We find then that
in the OPE between $e^{\mp \frac{i}{\sqrt{2}}X^3}$
and $G_{\N/SU(2)}$ the highest singularity must be\footnote{Note that
from this fact we cannot deduce, as in Section 2.2,
the existence of an $N=2$ structure
in $\N/SU(2)$, with R-current $i\sqrt{2} \d X^3$.}
$(z-w)^{-\half}$.
This result will be used in the next subsection.

The identification \iden, which can be stated more explicitly as
\eqn{idendos}{
e^{\pm i \sqrt{\frac1{2k}}Y \pm i\beta Z}=
e^{\mp \frac{i}2 H_3 \mp \frac{i}{\sqrt{2}}X^3}~,
}
has some consequences. From \stul, \jzeta, \stsutwo\ and \enedos\ we have
\eqn{katres}{
K^3=i\sqrt{2k}\d Y~.
}
Let $M^3$ be the $U(1)$ R-current of the worldsheet $N=2$ structure
obtained in the quotient $\N/U(1)$, where the $U(1)$ is generated by
the supercurrent $\chi^3+\theta K^3$. We have
\eqn{emetres}{
M^3=2i\beta \d Z~,
}
where $\beta$ and $Z$ are defined in \defbeta, \zeta, \zetazeta.
Equating the exponents in \idendos\
and using \bosfer\ and \jsuone\ it follows that
\eqn{idencor}{
M^3 = -\frac1{k}K^3 + k^3_{su(2)_2} + 2 k^3_{su(2)_1}~.
}
It was shown in  \cite{ags} that
if $\N$ contains a supersymmetric $SU(2)_{4k}$ affine symmetry
and has the properties required to yield spacetime $N=2$ supersymmetry,
then a {\it sufficient} condition to extend spacetime supersymmetry
to $N=3$ is that $\N$ contains a dimension-1 field
$k^3_{su(2)_1}=-{i\over\sqrt{2}}\d X^3$,
which commutes with $\chi^a$, and such that \idencor\ holds.
Here we see that the existence of the field $X^3$ with these properties
is also {\it necessary}.

Moreover, eq. \idencor\ was obtained by looking at the $N=2$
subalgebra \enedos,
but it must also hold for the other two $N=2$ subalgebras of
(\ref{uno}) -- (\ref{cinco}).
The general relation is then
\eqn{idencorgen}{
M^a = -\frac1{k}K^a + k^a_{su(2)_2} + 2 k^a_{su(2)_1}~,
}
where $M^a$ is the $U(1)$ R-current of the $N=2$ structure
obtained in the quotient $\N/U(1)$ (with the $U(1)$ generated by
$\chi^a+\theta K^a$), and $k^a_{su(2)_{1,2}}$
were defined in \boscurdos\ and \jsuone.
Of course, for $a=1,2$, in order to obtain \idencorgen\ explicitly
in the same way \idencor\ was obtained from \idendos, different
basis for the spin fields must be chosen.

An important consequence of \idencorgen\ is that
\eqn{emerota}{\eqalign
{
K^a(z)M^b(w) & \sim {i\eps^{ab}_{~~c}M^c(w) \over z-w}~,
\cr
M^d(z)\chi^e(w) & \sim {(1 - \frac1{k}) ~ i \eps^{de}_{~~f}
\chi^f \over z-w }~.
}}

\ni
Summing up, we have seen that the existence of $N=3$ superconformal symmetry
in the two dimensional theory dual to $AdS_3 \times \N$ implies that:
\begin{enumerate}
\item $\N$ contains an affine $SU(2)$ symmetry at level $4k$.
\item In the quotient of $\N$ by each one of the three $SU(2)$ supercurrents
$\chi^a + \theta K^a$,
the worldsheet supersymmetry is extended to $N=2$,
and the corresponding three $U(1)$ R-currents $M^a$ satisfy \emerota.
\end{enumerate}

\ni
Moreover, it is easy to see that these two conditions are also
{\it sufficient} to
yield $N=3$ spacetime supersymmetry, and are equivalent
to the conditions formulated in~\cite{ags}.

Note that the fact that the $SU(2)$ currents $K^a$ rotate the
three $U(1)$ R-currents $M^a$ implies that the $SU(2)$ part
is embedded in $\N$ as a nontrivial fibration over $\N/SU(2)$.
Namely, if $\N$ is a direct product --
$\N=SU(2)\times {\N\over SU(2)}$ -- then $K^a$ commute with the
contributions to $M^a$ coming from ${\N\over SU(2)}$, and thus cannot satisfy
\emerota.
The only exception to this is when ${\N\over SU(2)}$ is trivial, namely,
$\N=SU(2)$. This holds for $k=3/4$ $(k'=3)$, and the conditions
\emerota\ are indeed satisfied in this case.
This is a special case of a series of solutions
$\N=SU(3)_{k'}/U(1)$ \cite{ags}, because $SU(2)_3 \simeq SU(3)_3/U(1)$ as
SCFTs.

\subsection{Spacetime $N=1$ supersymmetry}
An interesting consequence of the structure found above is that
in every background which allows spacetime $N=3$ supersymmetry,
there is also a different GSO projection leading to precisely $N=1$ boundary
supersymmetry\footnote{Indeed, this was shown to be the case for the examples
studied in \cite{yamaguchi,ags}; here we argue that this is general.}.
This algebra is constructed from the singlet of the
$\mathbf{\half} \otimes \mathbf{\half}$ (discussed after eq. \spintodos),
and is generated by
{\setlength\arraycolsep{1pt}
\renewcommand{\arraystretch}{1.6}
\begin{equation}
\label{spinuno}
\begin{array}{lcl}
S_{\half}  & = & \frac12(   S_{[-,+,+,-]} - i S_{[-,-,-,+]})~,\\
S_{-\half} & = & \frac12 (  S_{[+,-,+,-]} - i S_{[+,+,-,+]})~.
\end{array}
\end{equation}}
\def\spinuno{(\ref{spinuno})}

\ni Note that the difference between these two operators and
$S^3_{\pm \half}$ in \spintodos\ is that
the sign of $H_2$ is changed (different GSO projection) and
the relative sign between the terms is now negative for the singlet.
Using the fact that in the OPE between $e^{\mp \frac{i}{\sqrt{2}}X^3}$
and $G_{\N/SU(2)}$, the highest singularity is $(z-w)^{-\half}$
(see the discussion after eq. \tresmedios),
the operators \spinuno\ can be checked to be BRST invariant.
Finally, they have regular OPEs with
$K^a$, as expected for the singlet.

\section{Spacetime small and large $N=4$ supersymmetry}
The analysis for these two cases is similar, mutatis mutandi, to that
of $N=2,3$, and we will only indicate the general arguments.

Small $N=4$ has an affine $SU(2)$ R-symmetry whose level is fixed
to be $k$ (the same as the $AdS_3$ level)
by the spacetime algebra. The central charge of $\N/SU(2)_{k}$ is then
\eqn{cargatc}{ c_{\N / SU(2)_{k}}=c_{\N}-\frac{3(k-2)}{k} - \frac32 =6~,}
and the construction of the spin fields leads to the
same supercharges constructed in \cite{gks}. As in Section 2,
the BRST condition shows that the worldsheet supersymmetry in
$\N/ SU(2)_{k}$ is extended to $N=2$, and from \cargatc\ this
actually means that there is a small $N=4$ supersymmetry in the
worldsheet\footnote{A $c=6$, $N=2$ SCFT
has necessarily small $N=4$ \cite{Banks:1988yz}.}.
The latter was realized in \cite{gks} by means of $\N / SU(2)_{k}=T^4$.

The analysis of the large $N=4$ case leads necessarily to
$\N=SU(2) \times   SU(2) \times  U(1)$~\cite{Elitzur:1998mm}.
The reason is that the spacetime R-symmetry requires the presence
of affine $SU(2)\times SU(2) \times U(1)$
in the worldsheet theory, and consistency of the spacetime algebra
requires the levels $k',k''$ of
the two $SU(2)$ models to satisfy $1/k = 1/k' + 1/k''$.
This implies that the
$(AdS_3)_k\times SU(2)_{k'}\times SU(2)_{k''}\times U(1)$
background is critical, and thus the unique one allowing large $N=4$
boundary supersymmetry.
This is compatible with the fact that the spin fields generating the Ramond
sector of the 10 free worldsheet fermions already have $\Delta=5/8$,
and hence the spacetime supercurrents must be constructed from them.
This is indeed the construction in \cite{Elitzur:1998mm}.
This model was also studied in
\cite{deBoer:1999rh},\cite{Figueroa-O'Farrill:2000ei}.

\section{Discussion}
In this work the necessary conditions have been
found for the internal CFT $\N$, imposed by the existence of $N=2,3,4$
SCFT in the boundary dual of string theory on $AdS_3 \times \N$.
Our results are summarized in Table 1.

\renewcommand{\arraystretch}{1.1}
\TABLE[h]{
 \begin{tabular}{|c|c|l|}\hline
 Boundary   & R-current    & Conditions on $\N$  \\\hline\hline
 $N=2$        & $U(1)$         & $\N \supset U(1)$ and  $\N/U(1)$ has $N=2$ SUSY.      \\ \hline
 $N=3$        &            & $\N \supset SU(2)_{4k}$ and  each $\N/K^a$ has         \\
 $( N=1)$   & $SU(2)$    &  $N=2$ SUSY with $U(1)$ R-currents  $M^a$  \\
              &            &   satisfying \emerota. \\ \hline
 Small $N=4$  & $SU(2)$   &  $\N = SU(2)_k \times M^4$                         \\ \hline
 Big $N=4$    & $SU(2)\!\! \times  \! SU(2)\!\! \times \! U(1)$  & $\cN=SU(2)_{k'}\!\! \times \! SU(2)_{k''}\!\! \times \! U(1), ~~
 1/k = 1/k' + 1/k''$
 \\ \hline
 \end{tabular}

\caption{This summarizes the necessary conditions on the internal CFT $\N$
imposed by different extended supersymmetries in the dual boundary CFT.
All the conditions are also sufficient.
In the $N=3$ case, $N=1$ is obtained for the other GSO projection;
$K^a$ are the generators of $SU(2)_{4k}$, and $\N/K^a$ denotes the
$\N/U(1)$ quotient SCFT with the $U(1)$ generated by the superfield
$\chi^a+\theta K^a$.
In small $N=4$,
$M^4$ is a $c=6$ unitary CFT with a small $N=4$ superconformal  symmetry.}
}

The case of {\em precisely} $N=1$ supersymmetry in the boundary
two dimensional CFT seems to be non-trivial.
Of course, it can be realized
as a byproduct of theories having higher
amounts of boundary supersymmetries. Examples of this
are the other GSO projection of theories
with boundary $N=3$ (see Section 3.3),
or a particular $Z_2$ orbifold of theories with $N=2$~\cite{GR}.
Recently, the work in ref. \cite{esy} implies that if the SCFT $\N$
contains a tri-critical Ising model, then $N=1$ supersymmetry can
be constructed in the boundary dual of $AdS_3$.
A new family of examples of this sort is $\N=SO(7)_{k'}/(G_2)_{k'+1}$,
which leads to precisely $N=1$ two dimensional supersymmetry in
spacetime\footnote{We thank Satoshi Yamaguchi for pointing out this result.}.


Finally, general properties of the spectrum of superstrings in
$AdS_3\times \N$, with extended supersymmetry, can be studied along
the lines of \cite{agstwo,agsthree}.

\section*{Acknowledgements}
We thank Jan de Boer, Shmuel Elitzur, Gast\'on Giribet,
Carmen Nu\~nez and Yaron Oz
for discussions and  Assaf Shomer for a
critical reading of an earlier draft of this work.
A.P.~is grateful to Instituto de
F\'{\i}sica de La Plata (Argentina) for hospitality during part of this work.
This work is supported in part by BSF --
American-Israel Bi-National Science Foundation, the Israel Academy
of Sciences and Humanities -- Centers of Excellence Program, the
German-Israel Bi-National Science Foundation, and the European RTN
network HPRN-CT-2000-00122.

\renewcommand{\baselinestretch}{0.87}

\newpage
\appendix
\renewcommand{\theequation}{\Alph{section}.\arabic{equation}}

\setcounter{equation}{0}
\section{Properties of the affine currents}

In this appendix we shall prove three
properties of the affine currents which we have used throughout the paper.
Let $\chi + \theta K$ be a generic worldsheet supercurrent
corresponding to the affine symmetries uplifted to the boundary theory, then
it holds for $\chi$ and $K$ that:
\begin{enumerate}
\item They are orthogonal.
\item They have the same normalization.
\item They commute with the three $SL(2,R)$ currents,
so they lie entirely in $\N$.

\end{enumerate}

\ni
Given the worldsheet supercurrent $G$,
remember that for any dimension-$\frac12$
superfield $\chi + \theta K$, we have
\eqn{opegmodos}{
\eqalign{
[G_s,K_m] & = -m \chi_{s+m}\,, \cr
\{G_s,\chi_n\} & = K_{s+n}\,.
}}
\ni
Property $1$ can be seen by
considering  the Jacobi identity
\eqn{supjacobi}{
\eqalign{
\{G_s,[\chi_t,K_q]\}-\{\chi_t,[K_q,G_s]\}+[K_q,\{G_s,\chi_t\}] & =0~, \cr
\{G_s,[\chi_t,K_q]\}-q\{\chi_t,\chi_{q+s}\}+[K_q,K_{s+t}] & =0~, \cr
\{G_s,[\chi_t,K_q]\} -q\delta_{t+q+s,0} + q\delta_{t+q+s,0} & =0~,
}}
for every set of $s,t,q$, and  it follows that
\eqn{jpzero}{
 K(z) \chi(w) \sim 0~.
}
Property 2 follows from
\eqn{mismanorma}
{\eqalign{
\{G_{-m-n},[\chi_n,K_m]\}-
\{\chi_n,[K_m,G_{-m-n}]\}+[K_m,\{G_{-m-n},\chi_n\}] & =0~, \cr
0 -m\{\chi_n,\chi_{-n}\}+ [K_m,K_{-m}]&=0~,
}}
for every $m$,$n$.

As for Property 3,
the commutators $\[L_{0,\pm1},R_0\]=0$, where $L_{0,\pm1}$ are given in
\stvir\ and $R_0\equiv \oint  K$,
show that there are no simple poles in the OPEs between $J^A$ and $K$.
A possible double pole is forbidden by Jacobi identities such as
\eqn{jacobijotas}{
\eqalign{
[K_n,[J^A_{-n+1},J^B_{-1}]] +
[J^A_{-n+1},[J^B_{-1},K_n]] + [J^B_{-1},[K_n,J^A_{-n+1}]] & = 0 ~, \cr
\ve^{ABC}\eta_{CD}[K_n,J^D_{-n}]+ 0 + 0  & = 0~,
}}
and from this  we find that
\eqn{opecero}{
J^A(z)K(w) \sim 0~.
}
Consider now the Jacobi identity,
\eqn{jacobi}{
\eqalign{
[G_s,[J^A_m,K_n]] + [J^A_m,[K_n,G_s]] + [K_n,[G_s,J^A_m]] & = 0~, \cr
0 + n[J^A_m,\chi_{n+s}]  -m [K_n,\psi^A_{m+s}]] & = 0 \,,
}}
where the first term is zero because of \opecero.
\ni
Choosing $n=0$ or $m=0$ in \jacobi, we have then
\eqn{ges}{
[K_0,\psi^A_t]  = [J^A_0,\chi_t] = 0~,
}
for every $t$. On the other hand, in the OPE, say,
between $K$ and $\psi^A$, the only terms that can
appear are
\eqn{opejf}{
K(z)\psi^A(w) \sim {\eta(w) \over z-w}~,
}
with $\Delta(\eta)= \frac12$, but from \ges,
\eqn{chicero}{
\eta(w)= [K_0, \psi^A(w)] = 0~.
}
In the same way we see that
\eqn{opecerom}{
\eqalign{
J^A(z)\chi(w) &\sim 0~, \cr
\psi^A(z)K(w) & \sim 0~.
}}

\ni Consider finally  Jacobi identities such as \eqn{superjacobi}{
\eqalign{
\{G_{-n+1},[\chi_n,J^A_{-1}]\}-\{\chi_n,[J^A_{-1},G_{-n+1}]\}
+[J^A_{-1},\{G_{-n+1},\chi_n\}] & =0~, \cr
0 +\{\chi_n,\psi^A_{-n}\}+ [J^A_{-1},K_1]&=0~, \cr 0
+\{\chi_n,\psi^A_{-n}\} + 0 &=0~, }}
where we have used \opecero\
and \opecerom. We conclude then that
\eqn{opecerof}{
\psi^A(z)\chi(w) \sim 0~, } and it follows from \opecero,
\opecerom\ and \opecerof\ that $\chi + \theta K$ commutes with
the $SL(2,R)$ sector of the worldsheet, and hence it lies entirely
in $\N$.

\setcounter{equation}{0}
\section{Realization of the symmetries for $N=3$}
In~\cite{ags} it was shown that eq. \idencor\ holds in case
$\N=SU(3)/U(1)$ and $\N=SO(5)/SO(3)$.
These backgrounds, as shown there explicitly,
yield spacetime $N=3$ (or $N=1$) supersymmetry.

In this appendix we will illustrate our results by seeing how the
geometric structure obtained in Section 3.2 is explicitly realized
in another background which was also shown to have $N=3$ spacetime
supersymmetry.

The background is $\N= \left( SU(2) \times SU(2) \times  U(1)
\right) / Z_2$, and was studied in~\cite{yamaguchi}, where the
explicit form of the spin fields is given. The two supersymmetric
$SU(2)$ WZW models have level $2k$. Let $\chi^a_{1,2} +
\theta K^a_{1,2}$ be the $SU(2)$ currents and $\lambda +\theta i \d
Y$ be the affine $U(1)$ current. The $Z_2$ orbifold acts as
\eqn{accionorbifold}{
\eqalign{
(K^a_1,K^a_2,Y) & \longrightarrow (K^a_2,K^a_1,-Y)~, \cr
(\chi^a_1, \chi^a_2, \lambda) & \longrightarrow (\chi^a_2, \chi^a_1, -\lambda)~.
}}
Define \eqn{sumaresta}{ \eqalign{ K^a_+ & =
K^a_1 + K^a_2 \qquad \chi^a_+ = \chi^a_1 + \chi^a_2 ~, \cr
K^a_- & = K^a_1 - K^a_2 \qquad \chi^a_- = \chi^a_1 - \chi^a_2~.
}}

\ni
The currents $\chi^a_+ + \theta K^a_+$ form a supersymmetric $SU(2)$ WZW model at level $4k$ which survives the
orbifold and is uplifted to the $SU(2)$ affine R-current of the $N=3$ superconformal algebra in spacetime.

The supercurrent  is
\eqn{superjapo}{
G_{\N}= \frac1k(\chi^a_1 k_{1a} - \frac{i}k \chi^1_1 \chi^2_1 \chi^3_1)
+ \frac1k(\chi^a_2 k_{2a} - \frac{i}k \chi^1_2 \chi^2_2 \chi^3_2)
+ i\lambda \d Y~.
}
Each superfield $\chi^a_+ + \theta K^a_+$
constitutes a supersymmetric $U(1)$ affine model with supercurrent
\eqn{superunojapo}{
G_{U^a(1)}= \frac1{2k} \chi^a_+ K^a_+~.
}
The coset of $\N$ by each of these $U(1)$
currents gives a theory whose supercurrent can be written as
\eqn{supercosetjapo}{
\eqalign{
G_{\N/U^3(1)}=G_{\N}-G_{U^3(1)}= & \frac1{2k}\chi_+^1k_+^1 +
\frac1{2k}\chi_+^2k_+^2 +
\cr
& \quad + \frac1{2k}\chi_-^1k_-^1
+ \frac1{2k}\chi_-^2k_-^2 + \frac1{2k}\chi_-^3 K_-^3
 + i\lambda \d Y~,
}} with the $a=1,2$ cases given by cyclic permutations. For each
$a$  there is an  $N=2$ structure which survives the orbifold,
whose $U(1)$ R-current  is
\eqn{errejapo} { M^a= -\frac1{k}K^a_+ -
\frac{i}{4k}\eps^{a}_{~bc}\chi^b_+ \chi^c_+ -
\frac{i}{4k}\eps^{a}_{~bc}\chi^b_- \chi^c_- -
\frac{i}{\sqrt{2k}}\chi^a_- \lambda~, }
and from \idencorgen\ we can
identify \eqn{idenequis}{2k^a_{su(2)_1}  =
-\frac{i}{4k}\eps^{a}_{~bc}\chi^b_- \chi^c_-
-\frac{i}{\sqrt{2k}}\chi^a_- \lambda~. }
The currents \errejapo\
clearly satisfy
\eqn{emerotajapo}{\eqalign{ K^a_+(z)M^b(w) & \sim
{i\eps^{ab}_{~~c}M^c(w) \over z-w}~, \cr
M^d(z)\chi_+^e(w) & \sim {(1 -
\frac1{k}) ~ i \eps^{de}_{~~f} \chi_+^f \over z-w }~, }} as expected from
\emerota.

\end{document}